\documentclass[aps,prb,10pt,reprint,twocolumn,final,amsfonts,amssymb,amsmath,hypertext,superscriptaddress]{revtex4-2}
\pdfoutput=1
\usepackage{xcolor,graphicx,hyperref}
\usepackage[inline]{enumitem}
\usepackage[version=3]{mhchem}
\usepackage{bm}
\usepackage{dcolumn}
\usepackage{physics}
\usepackage{siunitx}
\usepackage{subfigure}

\let\added\relax
\graphicspath{{plot}{figure}}

\begin{document}

\preprint{APS/123-QED}

\title{%
  Chirality-dependent second-order spin current\\
  in systems with time-reversal symmetry
}

\author{Ryosuke Hirakida}
\affiliation{Department of Physics, The University of Tokyo, Bunkyo, Tokyo 113-0033, Japan}

\author{Junji Fujimoto}
\affiliation{Department of Physics, The University of Tokyo, Bunkyo, Tokyo 113-0033, Japan}

\author{Masao Ogata}
\affiliation{Department of Physics, The University of Tokyo, Bunkyo, Tokyo 113-0033, Japan}
\affiliation{Trans-scale Quantum Science Institute, The University of Tokyo, Bunkyo, Tokyo 113-0033, Japan}

\date{\today}

\begin{abstract}
  Spin currents proportional to the first- and second-order of the electric field are calculated in a specific tight-binding model with time-reversal symmetry.
  Specifically, a tight-binding model with time-reversal symmetry is constructed with chiral hopping and spin-orbit coupling.
  The spin conductivity of the model is calculated using the Boltzmann equation.
  As a result, it is clarified that the first-order spin current of the electric field vanishes, while the second-order spin current can be finite.
  Furthermore, the spin current changes its sign by reversing the chirality of the model.
  The present results reveal the existence of spin currents in systems with time-reversal symmetry depending on the chirality of the system.
  They may provide useful information for understanding the chirality-dependent spin polarization phenomena in systems with time-reversal symmetry.
\end{abstract}

\maketitle

\section{Introduction}
\label{sec:introduction}
Three-dimensional structures that do not have mirror symmetry are called chiral.
This symmetry breaking gives rise to various physical properties such as electrical magnetochiral effect~\cite{Rikken2001} and chiral soliton lattice~\cite{Togawa2016}, which will be useful for spintronics or magnetic devices.
Among these properties, the spin polarization phenomenon called `chirality-induced spin selectivity' (CISS) has recently attracted much attention~\cite{Naaman2012, Naaman2015}.
CISS was first discovered in DNA in 2011~\cite{Gohler2011, Xie2011} and has been intensively studied in organic molecules.
CISS in DNA was reported to produce spin polarization rates of up to \SI{60}{\%}~\cite{Gohler2011}.
\added{The spin polarization observed here can be considered as a spin current, so CISS} is expected to be applied to efficient spin current generation techniques in the future.
In 2020, it was shown that the inorganic crystal \ce{CrNb3S6} also shows CISS~\cite{Inui2020, Nabei2020}.
More recently, CISS was confirmed also in inorganic crystals such as \ce{NbSi2} and \ce{TaSi2}~\cite{Shishido2021}.

While the theory of CISS in organic materials has been intensively investigated~\cite{Evers2022}, that in inorganic materials has not been explored.
Because the CISS is observed in the paramagnetic phase~\cite{Inui2020, Shishido2021}, the spin polarization occurs in the states with time-reversal symmetry.
However, it is known that a spin current linear in the electric field does not arise in single-channel systems with time-reversal symmetry~\cite{Bardarson2008}.
This theorem makes the construction of the CISS theory difficult.

In this study, \added{we assume that the spin currents are generated in the experiments of CISS and evaluate the spin conductivity in the constructed chiral tight-binding model with time-reversal symmetry.}
Specifically, we study a tight-binding model with chiral hopping and spin-orbit coupling to calculate the spin conductivity in the first- and second-order with respect to the electric field using the Boltzmann transport equation.
We show that the first-order spin conductivity vanishes in the tight-binding model, while the second-order spin conductivity can be finite.
Furthermore, the spin conductivity changes its sign by reversing the chirality of the model.
The present results reveal the existence of spin currents in systems with time-reversal symmetry and the existence of chirality-dependent spin currents.
They provide useful information for understanding the chirality-dependent spin polarization phenomena in systems with time-reversal symmetry.

This paper is organized as follows.
In Sect.~\ref{sec:symmetry-analysis}, we start with symmetry analysis of spin conductivities in the first- and second-order with respect to the electric field in the Hamiltonians with time-reversal symmetry.
In Sect.~\ref{sec:model-analysis}, we introduce a specific chiral tight-binding model and calculate the chirality dependence of the second-order spin conductivity.
We also discuss the parameters required for the chirality-dependent spin conductivity to arise.
Section~\ref{sec:conclusion} is devoted for discussions and conclusions.

\section{Symmetry Analysis}
\label{sec:symmetry-analysis}
We first discuss whether the spin current flows when time-reversal symmetry is imposed on the Hamiltonian in general.
We consider a model that is diagonal with respect to the spin but includes sublattice degrees of freedom, i.e., we assume that the spin $s_z$ ($s_z=\uparrow, \downarrow$) along the $z$-axis is a good quantum number.
In this case, the Hamiltonian in the wavenumber representation can be written as
\begin{align}
  \mathcal{H}(\vb*{k})=
  \begin{pmatrix}
    \mathcal{H}(\vb*{k}, \uparrow) & O \\
    O & \mathcal{H}(\vb*{k}, \downarrow) \\
  \end{pmatrix},
  \label{eq:gen-ham}
\end{align}
where $\mathcal{H}(\vb*{k}, \uparrow)$ and $\mathcal{H}(\vb*{k}, \downarrow)$ are matrices whose matrix elements are between the band index $n$.

By solving the Boltzmann transport equation within the approximation of constant relaxation time, the spin-resolved first-order conductivity $\sigma_{zz}(s_z)$ and spin-resolved second-order conductivity $\sigma_{zzz}(s_z)$ with respect to the electric field $E_z$, i.e., $j_z(s_z)=\sigma_{zz}(s_z)E_z+\sigma_{zzz}(s_z){E_z}^2+\cdots$, can be written as~\cite{Sodemann2015}
\begin{align}
  \sigma_{zz}(s_z)&=
  \frac{e^2\tau}{\hbar^2}\sum_n
  \int_{\mathrm{BZ}}\frac{\dd^3k}{(2\pi)^3}\
  \frac{\partial^2\varepsilon_{n}(\vb*{k},s_z)}{{\partial k_z}^2}
  f(\varepsilon_n(\vb*{k},s_z)),
  \label{eq:1st-order-cond}\\
  \sigma_{zzz}(s_z)&=
  \frac{e^3\tau^2}{\hbar^3}\sum_n
  \int_{\mathrm{BZ}}\frac{\dd^3k}{(2\pi)^3}\
  \frac{\partial^3\varepsilon_{n}(\vb*{k},s_z)}{{\partial k_z}^3}
  f(\varepsilon_n(\vb*{k},s_z)),
  \label{eq:2nd-order-cond}
\end{align}
where $e(<0)$ is the electric charge, $\tau$ represents the relaxation time, $\varepsilon_n(\vb*{k},s_z)$ represents the eigenenergy of the band index $n$ with wave number $\vb*{k}$ and spin $s_z$, and $f(\varepsilon)$ represents the Fermi distribution function.
$\int_{\mathrm{BZ}}\dd^3k$ means the integration over the entire Brillouin zone.
The first-order spin conductivity $\sigma^\mathrm{s}_{zz}$ and the second-order spin conductivity $\sigma^\mathrm{s}_{zzz}$ are defined as
\begin{align}
  \sigma^\mathrm{s}_{zz}&=\sigma_{zz}(\uparrow)-\sigma_{zz}(\downarrow),\\
  \sigma^\mathrm{s}_{zzz}&=\sigma_{zzz}(\uparrow)-\sigma_{zzz}(\downarrow).
\end{align}
We further suppose that the Hamiltonian in Eq.~\eqref{eq:gen-ham} has the time-reversal symmetry that can be written as $\Theta=-i s_y K$, where $K$ is the complex conjugate operator.
In this case, the eigenenergy of the Hamiltonian in Eq.~\eqref{eq:gen-ham} satisfies
\begin{align}
  \varepsilon_n(\vb*{k}, s_z)=\varepsilon_n(-\vb*{k}, -s_z).
  \label{eq:sym-energy}
\end{align}
Using this equation in Eqs.~\eqref{eq:1st-order-cond} and \eqref{eq:2nd-order-cond}, we can readily see that
\begin{align}
  \sigma_{zz}(s_z)&=\sigma_{zz}(-s_z),\label{eq:sym-first-cond}\\
  \sigma_{zzz}(s_z)&=-\sigma_{zzz}(-s_z),\label{eq:sym-second-cond}
\end{align}
which means that the first-order spin conductivity $\sigma^\mathrm{s}_{zz}$ vanishes in systems satisfying the above time reversal symmetry, while the second-order spin conductivity $\sigma^\mathrm{s}_{zzz}$ can be finite even under such conditions.
Note that this argument does not hold if the degrees of freedom of the time-reversal operator except for spin, for example, the sublattice degrees of freedom are not an identity matrix.

\added{One of the possible interpretations of this result is as follows. For the first-order spin conductivity $\sigma^s_{zz}$ to be finite, it is usually necessary for the magnetization $M_z$ be incorporated into the system. The magnetization $M_z$ is zero in the present system because the system has time-reversal symmetry, and thus the first-order spin conductivity $\sigma^s_{zz}$ vanishes.
However, the system has a finite orbital magnetization $M_z$ proportional to the electric field $E_z$ in the current-carrying state. This magnetization is caused by the conduction electrons taking a circular motion in the xy-plane due to the chiral hopping. Therefore, the current-carrying state breaks the time-reversal symmetry in the sense that it has a finite magnetization $M_z$. Now, if we construct a response that is itself linear in $E_z$ with respect to the current-carrying state, we may obtain a finite spin current. The process is facilitated by the spin-orbit coupling. As a result, the second-order spin conductivity $\sigma^s_{zzz}$ becomes finite.}

\added{This result is consistent with Bardarson's theorem~\cite{Bardarson2008} discussed in Sec.~\ref{sec:introduction} because Bardarson shows that first-order spin currents vanish under time-reversal symmetry and does not mention second-order spin currents. Several papers show that first-order spin currents can appear by relaxing the assumptions of Bardarson's theorem. For example, Utsumi et al. showed that first-order spin currents appear when each site of a helical molecule has two orbitals~\cite{Utsumi2020}. This result is also consistent with that of the present paper because we show that the linear spin current vanishes under the assumption that the time-reversal operator does not exchange the sublattice degrees of freedom. The model used in Ref.~\cite{Utsumi2020} does not satisfy this assumption.}

\section{Model Analysis}
\label{sec:model-analysis}
\begin{figure}[tbp]
  \centering%
  \subfigure[]{\includegraphics[height=.30\columnwidth,page=0]{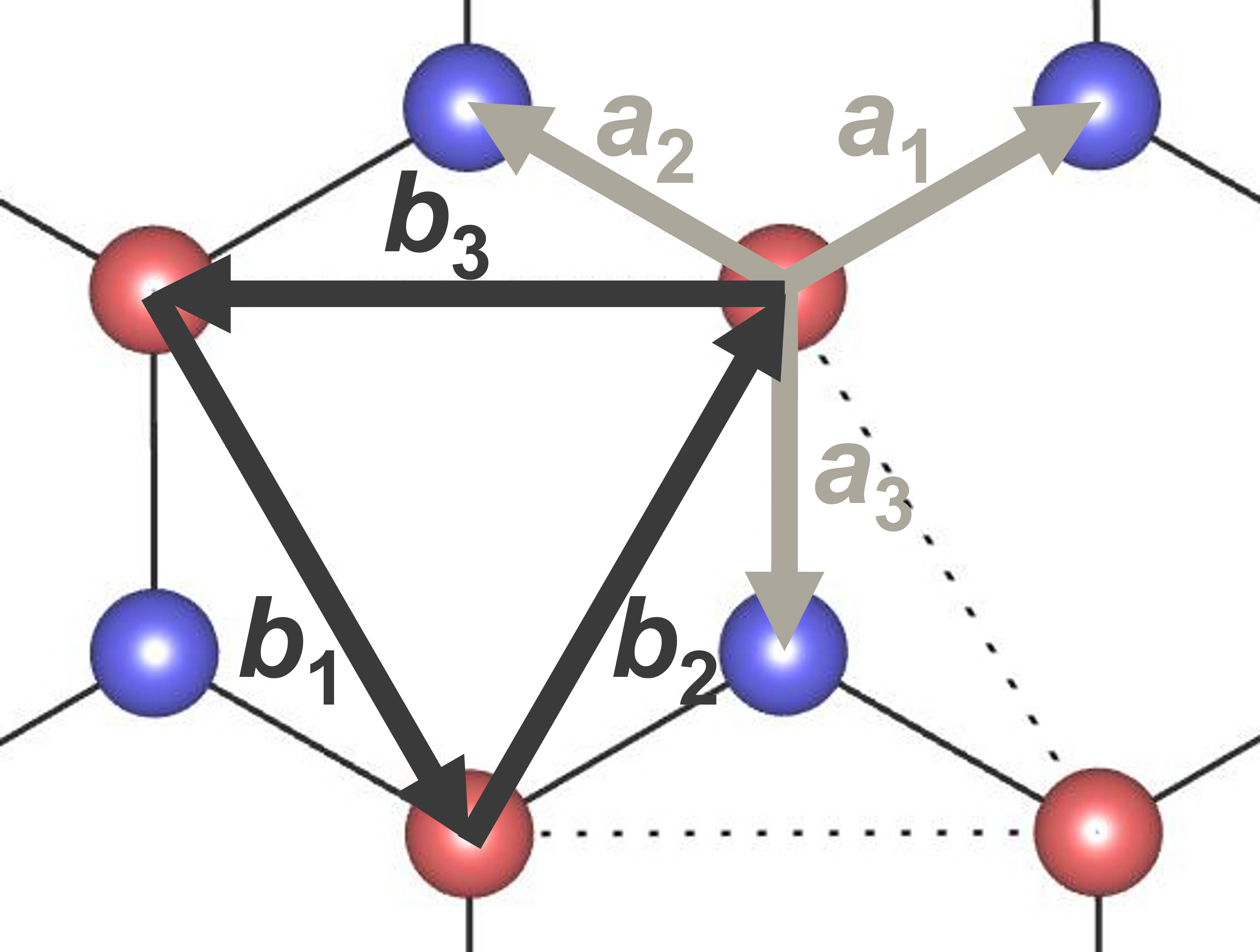}\label{sub:model-lattice-vectors}}\hspace{.02\columnwidth}%
  \subfigure[]{\includegraphics[height=.30\columnwidth,page=0]{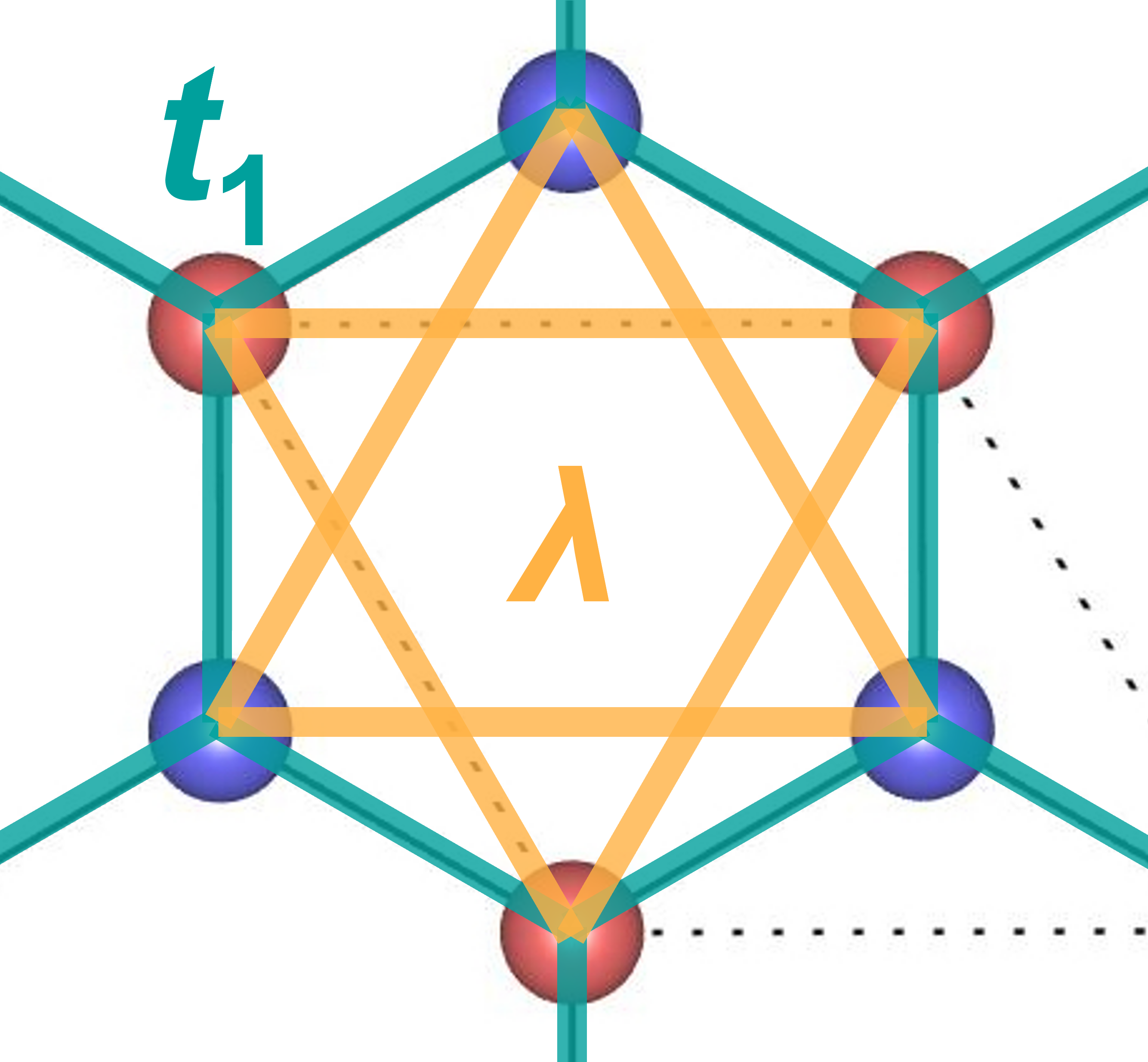}\label{sub:model-hopping}}\\%
  \subfigure[]{\includegraphics[width=.39\columnwidth,page=0]{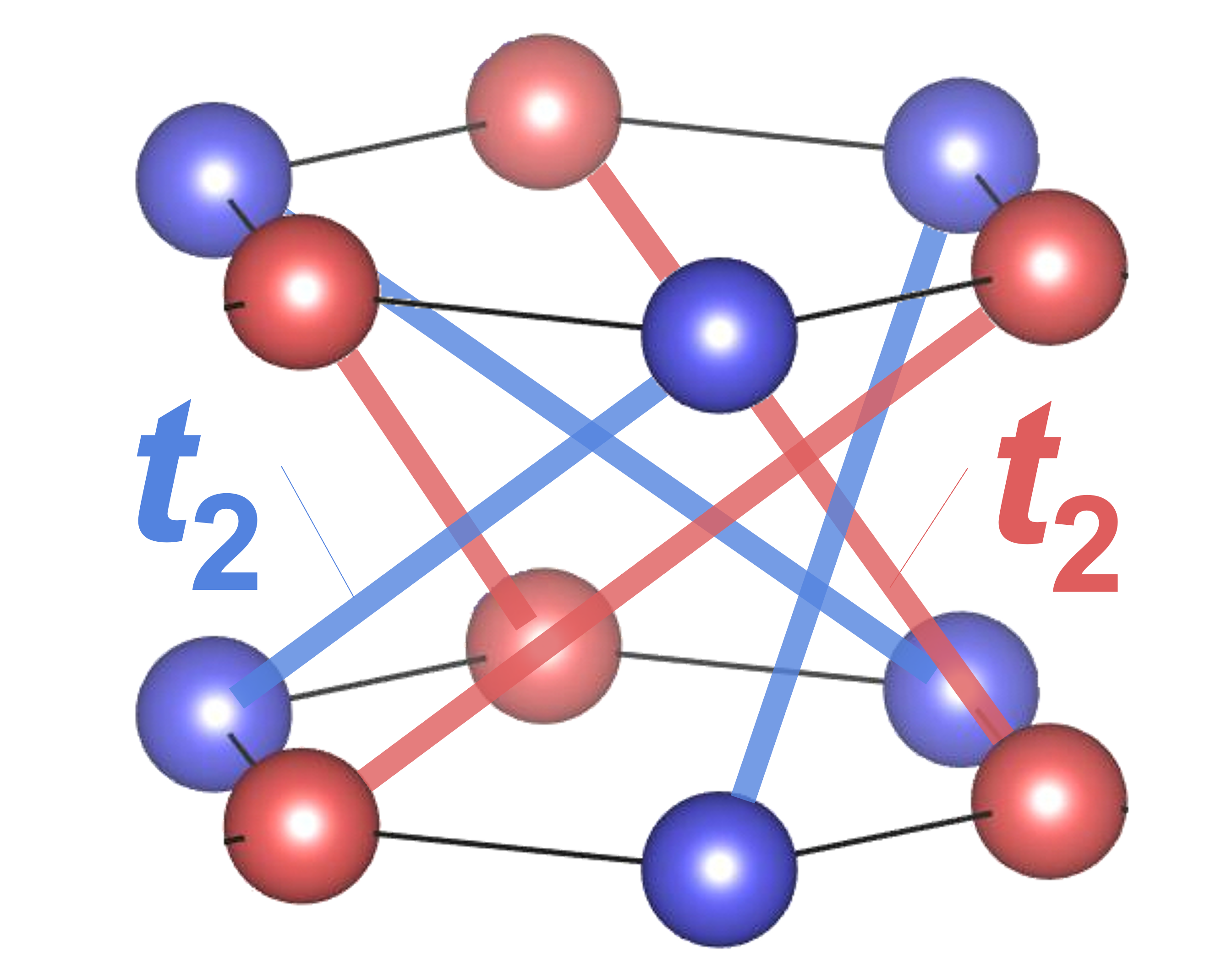}\label{sub:model-right-handed}}%
  \subfigure[]{\includegraphics[width=.39\columnwidth,page=0]{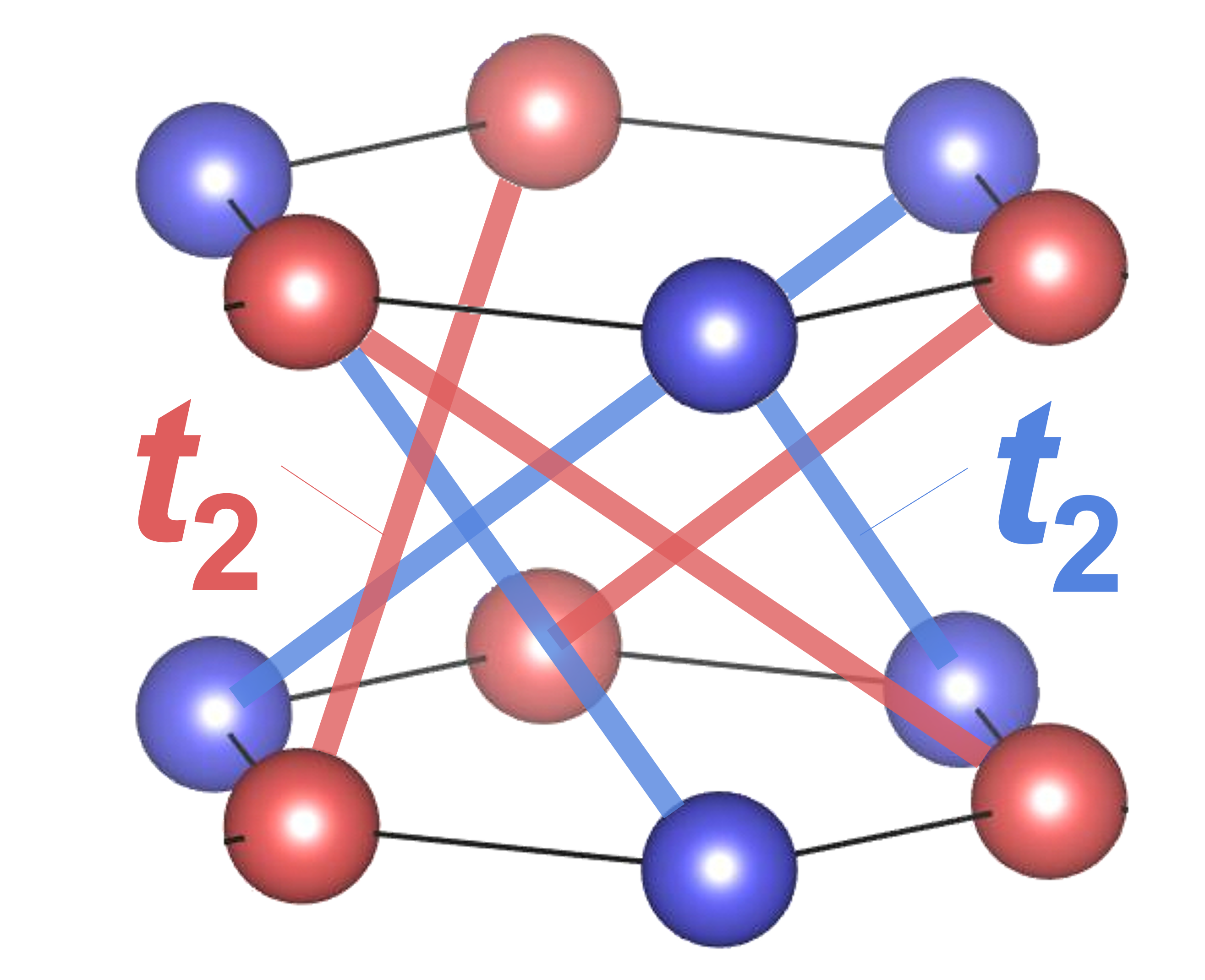}\label{sub:model-left-handed}}%
  \caption{%
    Tight-binding model described by the Hamiltonian in Eq.~\eqref{eq:hamiltonian}~\cite{Yoda2015}.
    \subref{sub:model-lattice-vectors} $xy$-plane of the present model. The vectors $\vb*{a}_a$ and $\vb*{b}_a$ ($a=1,2,3$) are shown in the figure.
    \subref{sub:model-hopping} Nearest-neighbor hopping (green) and next nearest-neighbor spin-orbit coupling (yellow) in the $xy$ plane.
    \subref{sub:model-right-handed}\subref{sub:model-left-handed} Chiral hopping between the nearest-negihbor layers: \subref{sub:model-right-handed} the right-handed model ($\chi=+1$) and \subref{sub:model-left-handed} the left-handed model ($\chi=-1$).
  }%
  \label{fig:model}
\end{figure}

Next, we discuss the chirality dependence of the second-order spin conductivity in a specific tight-binding model\added{, which captures the essence of chiral crystals}.
We introduce a three-dimensional tight-binding model with chiral hopping shown in Fig.~\ref{fig:model}~\cite{Yoda2015}.
This model consists of two-dimensional honeycomb lattices stacked along the $z$-axis.
The Hamiltonian can be written as
\begin{widetext}
  \begin{align} 
    H&=H_\mathrm{hop}+H_\mathrm{chiral}+H_\mathrm{SOC},\label{eq:hamiltonian}\\
    H_\mathrm{hop}&=
    t_1\sum_{\langle i,j\rangle,l,\alpha}c_{i,l,\alpha}^\dagger c_{j,l,\alpha}+\mathrm{h.c.},\\
    H_\mathrm{chiral}&=
    t_2
    \left(
      \sum_{i\in\mathrm{A},l,\alpha}
      \sum_{a=1}^{3}
      c_{i+\chi\vb*{b}_a,l+1,\alpha}^\dagger c_{i,l,\alpha}
      +\sum_{i\in\mathrm{B},l,\alpha}
      \sum_{a=1}^{3}
      c_{i-\chi\vb*{b}_a,l+1,\alpha}^\dagger c_{i,l,\alpha}
    \right)
    +\mathrm{h.c.},\\
    H_\mathrm{SOC}&=
    i\lambda
    \left(
      \sum_{i\in\mathrm{A},l,\alpha,\beta}
      \sum_{a=1}^{3}(
        c_{i+\vb*{b}_a,l,\alpha}^\dagger (s_z)_{\alpha\beta} c_{i,l,\beta}
        -c_{i-\vb*{b}_a,l,\alpha}^\dagger (s_z)_{\alpha\beta} c_{i,l,\beta}
      )
    \right.\nonumber\\
    &\left.
      +\sum_{i\in\mathrm{B},l,\alpha,\beta}
      \sum_{a=1}^{3}(
        c_{i-\vb*{b}_a,l,\alpha}^\dagger (s_z)_{\alpha\beta} c_{i,l,\beta}
        -c_{i+\vb*{b}_a,l,\alpha}^\dagger (s_z)_{\alpha\beta} c_{i,l,\beta}
      )
    \right)
    +\mathrm{h.c.},
  \end{align}
\end{widetext}
where $c_{i,l,\alpha}$ ($c_{i,l,\alpha}^\dagger$) is the annihilation (creation) operator of an electron with spin $\alpha$ at the site $i$ in the $xy$-plane of the $l$-th layer.
$H_\mathrm{hop}$ denotes the nearest-neighbor hopping in the $xy$-plane with the transfer integral $t_1\in \mathbb{R}$.
The $\langle i,j\rangle$ spans the nearest-neighbor sites of the honeycomb lattices as shown in Fig.~\ref{fig:model}\subref{sub:model-hopping}.
$H_\mathrm{chiral}$ denotes the chiral hopping between the nearest-neighbor layers with the transfer integral $t_2\in \mathbb{R}$.
$\vb*{a}_a\ (a=1,2,3)$ are the vectors pointing the nearest-neighbor sites in the same plane, and $\vb*{b}_a\ (a=1,2,3)$ are the vectors pointing the next-nearest-neighbor sites in the same plane.
$\sum_{i\in\mathrm{A}}$ ($\sum_{i\in\mathrm{B}}$) means the summation over the sites in the A (B) sublattice.
$\chi=\pm 1$ is a parameter which represents the chirality of the model.
As we can see from Fig.~\ref{fig:model}\subref{sub:model-lattice-vectors}, when we change the sign of $\chi$, the chirality of the hopping is reversed.
Figures~\ref{fig:model}\subref{sub:model-right-handed} and \subref{sub:model-left-handed} show the right-handed and left-handed models, respectively.
$H_\mathrm{SOC}$ represents the spin-orbit coupling associated with the next-nearest-neighbor hopping in the layer propotional to $s_z$ with the coefficient $\lambda\in \mathbb{R}$.
$(s_z)_{\alpha\beta}$ is the $(\alpha,\beta)$ component of the spin operator along the $z$-axis.

\begin{figure}[tbp]
  \centering%
  \includegraphics[width=.70\columnwidth,page=1]{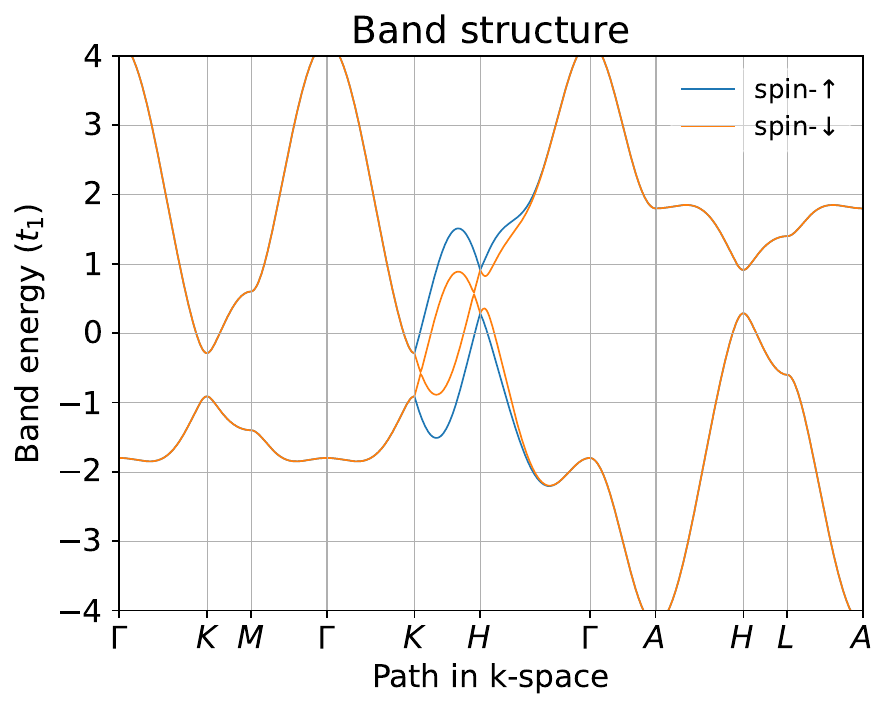}
  \caption{Band structure of the present model. The parameters are $t_2=0.2t_1$ and $\lambda=0.06t_1$.}
  \label{fig:band-structure}
\end{figure}

The Hamiltonian in the wavenumber representation can be written as
\begin{align} 
  \mathcal{H}(\vb*{k},s_z,\chi)
  &=d_0(\vb*{k})I+d_x(\vb*{k})\tau_x\nonumber\\
  &+d_y(\vb*{k})\tau_y+d_z(\vb*{k},s_z,\chi)\tau_z,\label{eq:wave-ham}\\
  d_0(\vb*{k})
  &=2t_2\sum_{a=1}^3\cos(\vb*{k}\cdot\vb*{b}_a)\cos(k_zc),\\
  d_x(\vb*{k})
  &=t_1\sum_{a=1}^3\cos(\vb*{k}\cdot\vb*{a}_a),\\
  d_y(\vb*{k})
  &=t_1\sum_{a=1}^3\sin(\vb*{k}\cdot\vb*{a}_a),\\
  d_z(\vb*{k},s_z,\chi)
  &=-2\sum_{a=1}^3\sin(\vb*{k}\cdot\vb*{b}_a)(\chi t_2\sin(k_zc)+2s_z\lambda),\label{eq:wave-ham-dz}
\end{align}
where $I$ is the identity matirx of size 2, $\vb*{\tau}=(\tau_x, \tau_y, \tau_z)$ are the Pauli matrices representing the sublattice degrees of freedom.
The eigenenergies of the Hamiltonian in Eq.~\eqref{eq:wave-ham} are expressed as
\begin{align} 
  &\varepsilon_\pm(\vb*{k},s_z,\chi)\nonumber\\
  =&d_0(\vb*{k})\pm\sqrt{{d_x(\vb*{k})}^2+{d_y(\vb*{k})}^2+{d_z(\vb*{k},s_z,\chi)}^2}.\label{eq:wave-ham-eigenenergy}
\end{align}
The band structure of this model is shown in Fig.~\ref{fig:band-structure}.
Apparently, $\varepsilon_\pm(\vb*{k},s_z,\chi)$ satisfies the time reversal condition in Eq.~\eqref{eq:sym-energy}, and thus the relation in Eq.~\eqref{eq:sym-first-cond} holds, i.e., the first-order spin conductivity $\sigma^\mathrm{s}_{zz}$ vanishes.

To study the chirality dependence of the second-order spin conductivity $\sigma^\mathrm{s}_{zzz}(\chi)$, we study the dependence of $\varepsilon_\pm(\vb*{k},s_z,\chi)$ on $\chi$.
From Eq.~\eqref{eq:wave-ham-dz}, we obtain
\begin{align} 
  d_z(\vb*{k},s_z,\chi)&=-d_z(\vb*{k},-s_z,-\chi).\label{eq:chi-dz}
\end{align}
Then we can readily see that
\begin{align}
  \varepsilon_\pm(\vb*{k},s_z,\chi)&=\varepsilon_\pm(\vb*{k},-s_z,-\chi),\\
  \frac{\partial^3\varepsilon_{n}(\vb*{k},s_z,\chi)}{{\partial k_z}^3}&=-\frac{\partial^3\varepsilon_{n}(\vb*{k},-s_z,-\chi)}{{\partial k_z}^3}.
\end{align}
Considering these equations with Eqs.~\eqref{eq:2nd-order-cond} and \eqref{eq:wave-ham-eigenenergy}, we obtain
\begin{align} 
  \sigma_{zzz}(s_z,\chi)&=\sigma_{zzz}(-s_z,-\chi).
\end{align}
Therefore, we can see
\begin{align} 
  \sigma_{zzz}^{\mathrm{s}}(\chi)&=-\sigma_{zzz}^{\mathrm{s}}(-\chi),
\end{align}
which means that the spin conductivity of the right-handed model $\sigma_{zzz}^{\mathrm{s}}(+1)$ and that of the left-handed model $\sigma_{zzz}^{\mathrm{s}}(-1)$ are of opposite signs in the present chiral tight-binding model.

In the following, we show that the chirality and the spin-orbit coupling of the present model play crucial roles in the chirality-dependent second-order spin current.
Here we fix the chirality of the model $\chi$.
When we assume that $t_2=0$, i.e., the model has no chirality, we obtain
\begin{align}
  \sigma_{zzz}(s_z,\chi)=\sigma_{zzz}(-s_z,\chi),
\end{align}
because $d_z(\vb*{k},s_z,\chi)=-d_z(\vb*{k},-s_z,\chi)$ holds as seen from Eq.~\eqref{eq:wave-ham-dz}.
This means that the second-order spin conductivity $\sigma^\mathrm{s}_{zzz}(\chi)$ vanishes under $t_2=0$.
Next, when we assume that $\lambda=0$, i.e., the model has no spin-orbit coupling, we obtain
\begin{align}
  \sigma_{zzz}(s_z,\chi)=\sigma_{zzz}(-s_z,\chi),
\end{align}
because $d_z(\vb*{k},s_z,\chi)=d_z(\vb*{k},-s_z,\chi)$ holds from Eq.~\eqref{eq:wave-ham-dz}.
This means that the second-order spin conductivity $\sigma^\mathrm{s}_{zzz}(\chi)$ again vanishes under $\lambda=0$.
From the above discussion, we can conclude that the chirality and the spin-orbit coupling are necessary for the chirality-dependent second-order spin current in the present model.

\begin{figure}[tbp]
  \centering%
  \includegraphics[width=.70\columnwidth,page=1]{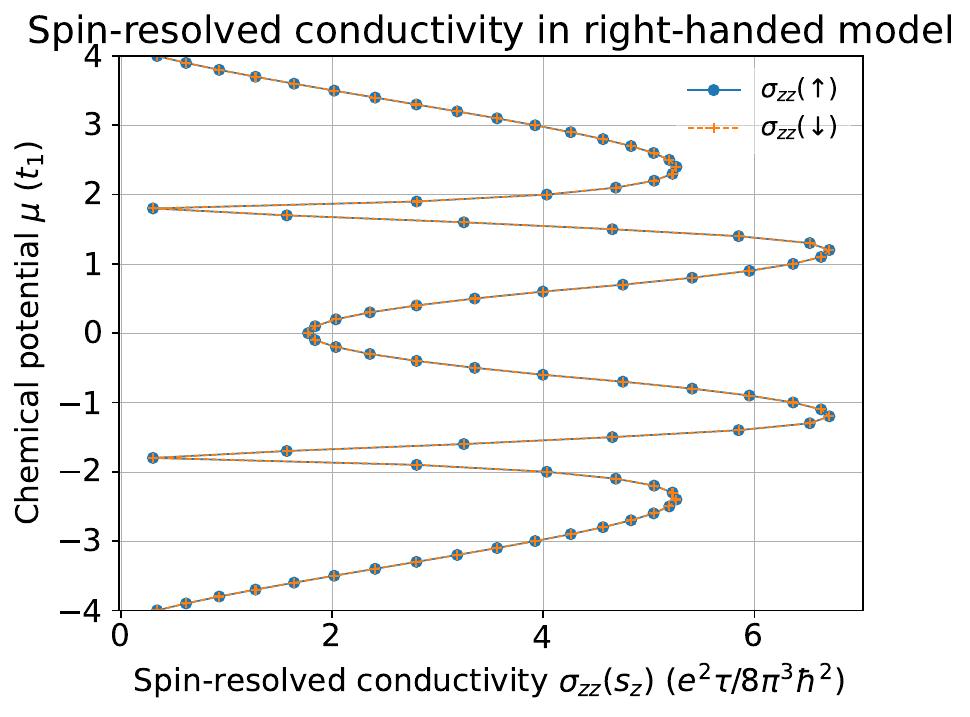}%
  \caption{
    Calculated spin-resolved first-order conductivity in the right-handed model for $\uparrow$-spin (blue, solid line) and $\downarrow$-spin (orange, dashed line). The parameters are $t_2=0.2t_1$ and $\lambda=0.06t_1$. The temperature $T$ is set at zero.
  }%
  \label{fig:calc-first}
\end{figure}

\begin{figure}[tbp]
  \centering%
  \includegraphics[width=.70\columnwidth,page=1]{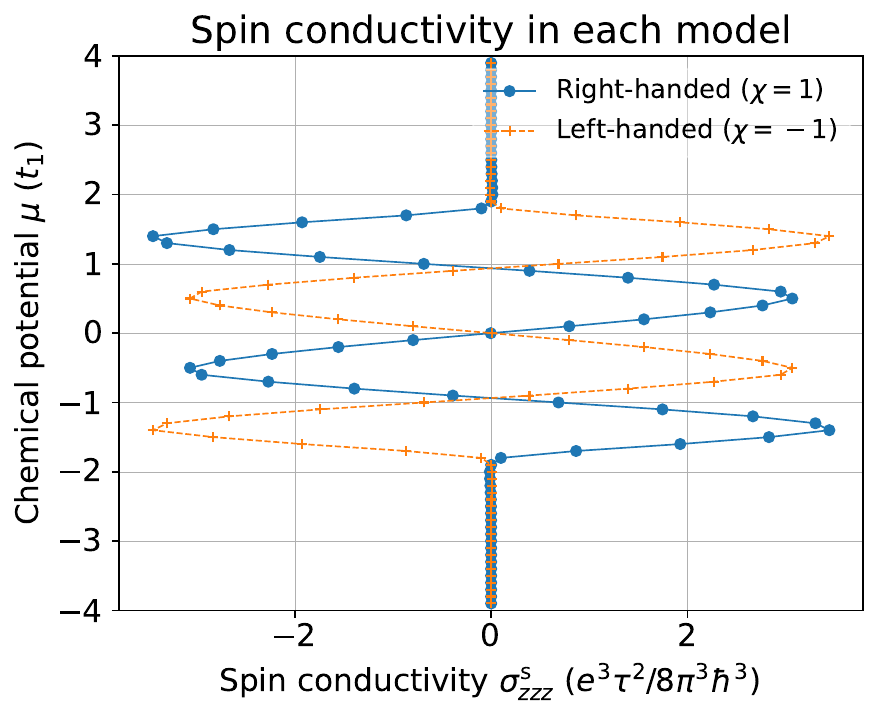}%
  \caption{
    Calculated second-order spin conductivity for the right-handed model (blue, solid line) and left-handed model (orange, dashed line). The parameters and temperature are the same as in Fig.~\ref{fig:calc-first}.
  }%
  \label{fig:calc-second}
\end{figure}

The calculated spin-resolved first-order conductivities and second-order spin conductivities are shown in Figs.~\ref{fig:calc-first} and \ref{fig:calc-second}.
Figure~\ref{fig:calc-first} depicts the chemical potential dependences of the spin-resolved first-order conductivities.
The spin-$\uparrow$ conductivity has the same dependence as that of the spin-$\downarrow$ conductivity, which results in the zero spin conductivity in the first order with respect to the electric field, as discussed in Sect.~\ref{sec:symmetry-analysis}.
Figure~\ref{fig:calc-second} depicts the chemical potential dependences of the second-order spin conductivities.
The spin conductivity in the left-handed model has the opposite sign as compared with that in the right-handed model, indicating that the sign of the spin conductivity can be controlled by reversing the chirality, as discussed above.
Figure~\ref{fig:calc-second} shows that the second-order spin conductivity rapidly changes its sign depending on the chemical potential in the region of $-2<\mu/t_1<2$.
This will originate from the difference in the energy dispersions of the spin-$\uparrow$ and spin-$\downarrow$ in the region of $-2<\varepsilon/t_1<2$ as shown in Fig.~\ref{fig:band-structure}.

\section{Conclusion}
\label{sec:conclusion}
We explored the existence of spin current and its chirality dependence in systems with time-reversal symmetry.
We evaluated the spin conductivity in the system with time-reversal symmetry based on the Boltzmann transport equation.
We found that the first-order spin conductivity with respect to the electric field vanishes, while the second-order spin conductivity can be finite.
We also evaluated the second-order spin conductivity using a specific chiral tight-binding model.
As a result, we revealed the existence of the second-order spin conductivity depending on the chirality of the system.
We also found that the chirality and the spin-orbit coupling play crucial roles in the chirality-dependent spin current\added{, and this would also hold for general systems}.

We considered in this study the second-order spin conductivity with respect to the electric field because the first-order spin conductivity vanishes for the non-interacting single-channel systems with time-reversal symmetry~\cite{Bardarson2008}.
However, spin polarization phenomena in the systems with spin hybridization have not been considered yet.
The role of weak antisymmetric spin-orbit coupling in spin polarization phenomena is also discussed~\cite{Inui2020}.
The mechanism in which the spin-flipping components of spin-orbit couplings affect the spin-polarized current remains to be investigated in future studies.

\added{In this paper, we assumed the relaxation time approximation, which does not correctly capture the effect of multi-band scattering. When we take into account the multi-band scattering, the magnitude of the spin conductivity can be changed. However, we think that the conclusion of this study does not change even if the multi-band scattering is considered. The fact that the first-order conductivity $\sigma_{zz}$ vanishes is protected by symmetry, while the conclusion that the second-order conductivity $\sigma_{zzz}$ is finite remains the same.}

\added{Furthermore, we did not discuss the possible contributions from the anomalous velocity~\cite{Nagaosa2010}, which will be an interesting problem.
However, we believe that the contribution of the anomalous velocity in $\sigma_{zzz}$ is not the leading term in the lifetime $\tau$ and that the leading term is determined from the Boltzmann transport theory as discussed in Ref.~\cite{Sodemann2015}.
No previous study considers such a contribution exactly, and it remains unclear whether the contribution is significant or not.
We believe that such a consideration should be made in future studies of this paper.}
\section{Acknowledgement}
\label{sec:acknowledgement}
This work was supported by Grants-in-Aid for Scientific Research from the Japan Society for the Promotion of Science (Grant No. JP18H01162), and by JST-Mirai Program (Grant No. JPMJMI19A1).
R.~H. was supported by Iwadare Scholarship Foundation and the Program for Leading Graduate Schools (MERIT-WINGS).

\bibliographystyle{apsrev4-2}
\bibliography{main.bbl}

\end{document}